\documentclass[preprint2]{aastex63}

\def\gsim{\mathrel{\rlap{\lower 4pt \hbox{\hskip 1pt $\sim$}}\raise 1pt
\hbox {$>$}}}
\def\lsim{\mathrel{\rlap{\lower 4pt \hbox{\hskip 1pt $\sim$}}\raise 1pt
\hbox {$<$}}}

\received{}
\revised{}
\accepted{}
\shorttitle{Resurrection of SN 2018ivc}
\shortauthors{Maeda et al.}
\graphicspath{{./}{figures/}}

\begin{document}

\title{Resurrection of type IIL supernova 2018ivc: 
Implications for a binary evolution sequence connecting hydrogen-rich and -poor progenitors}

\correspondingauthor{Keiichi Maeda}
\email{keiichi.maeda@kusastro.kyoto-u.ac.jp}

\author[0000-0003-2611-7269]{Keiichi Maeda}
\affiliation{Department of Astronomy, Kyoto University, Kitashirakawa-Oiwake-cho, Sakyo-ku, Kyoto, 606-8502. Japan}

\author[0000-0003-2475-7983]{Tomonari Michiyama}
\affiliation{Department of Earth and Space Science, Osaka University, 1-1 Machikaneyama, Toyonaka, Osaka 560-0043, Japan}
\affiliation{National Astronomical Observatory of Japan, National Institutes of Natural Sciences (NINS), 2-21-1 Osawa, Mitaka, Tokyo 181-8588, Japan}

\author[0000-0002-0844-6563]{Poonam Chandra}
\affiliation{National Radio Astronomy Observatory, 520 Edgemont Road,
Charlottesville VA 22903, USA}

\author[0000-0003-4501-8100]{Stuart Ryder}
\affiliation{School of Mathematical and Physical Sciences, Macquarie University, NSW 2109, Australia}
\affiliation{Macquarie University Research Centre for Astronomy, Astrophysics \& Astrophotonics, Sydney, NSW 2109, Australia}

\author[0000-0002-1132-1366]{Hanindyo Kuncarayakti}
\affiliation{Tuorla Observatory, Department of Physics and Astronomy, FI-20014 University of Turku, Finland} 
\affiliation{Finnish Centre for Astronomy with ESO (FINCA), FI-20014 University of Turku, Finland}

\author[0000-0002-1125-9187]{Daichi Hiramatsu}
\affiliation{Center for Astrophysics \textbar{} Harvard \& Smithsonian, 60 Garden Street, Cambridge, MA 02138-1516, USA}
\affiliation{The NSF AI Institute for Artificial Intelligence and Fundamental Interactions}
\affiliation{Las Cumbres Observatory, 6740 Cortona Drive, Suite 102, Goleta, CA 93117-5575, USA}
\affiliation{Department of Physics, University of California, Santa Barbara, CA 93106-9530, USA}

\author[0000-0001-6186-8792]{Masatoshi Imanishi}
\affiliation{National Astronomical Observatory of Japan, National Institutes of Natural Sciences (NINS), 2-21-1 Osawa, Mitaka,
Tokyo 181-8588, Japan}
\affiliation{Department of Astronomy, School of Science, The Graduate University for Advanced Studies, SOKENDAI, Mitaka, Tokyo 181-8588, Japan}



\begin{abstract}
Long-term observations of synchrotron emission from supernovae (SNe), covering more than a year after the explosion, provide a unique opportunity to study the poorly-understood evolution of massive stars in the final millennium of their lives via changes in the mass-loss rate. Here, we present a result of our long-term monitoring of a peculiar type IIL SN 2018ivc, using the Atacama Large Millimeter/submillimeter Array (ALMA). Following the initial decay, it showed unprecedented rebrightening starting at $\sim$ a year after the explosion. This is one of the rare examples showing such rebrightening in the synchrotron emission, and the first case at millimeter wavelengths. We find it to be in the optically-thin regime unlike the optically-thick centimeter emission. As such, we can robustly reconstruct the distribution of the circumstellar matter (CSM) and thus the mass-loss history in the final $\gsim 1,000$ years. We find that the progenitor of SN 2018ivc had experienced a very high mass-loss rate ($\gsim 10^{-3} M_\odot$ yr$^{-1}$) $\sim 1,500$ years before the explosion, which was followed by a moderately high mass-loss rate ($\gsim 10^{-4} M_\odot$ yr$^{-1}$) up until the explosion. From this behavior, we suggest SN 2018ivc represents an extreme version of a binary evolution toward SNe IIb, which bridges the hydrogen-poor SNe (toward SNe Ib/c, without a hydrogen envelope) and hydrogen-rich SNe (SNe IIP, with a massive envelope). 
\end{abstract}



\section{Introduction} \label{sec:intro}

\begin{deluxetable*}{cccccccc}
\tablecaption{ALMA measurements of SN 2018ivc}
\tablewidth{0pt}
\tablehead{
\colhead{MJD} & \colhead{Phase$^{\rm a}$} & \colhead{Frequency$^{\rm b}$} & \colhead{$F_{\nu}^{\rm c}$} & Array & \colhead{Resolution$^{\rm d}$} & \colhead{ID} & \colhead{PI}\\
\colhead{} & \colhead{(Days)} & \colhead{(GHz)} & \colhead{(mJy)} & \colhead{} & \colhead{} & \colhead{} & \colhead{}
}
\startdata
{\bf Band 3} & & & & & & & \\\hline
58449.1$^{\rm e}$  &    4.1 & 100.0 & $4.25  \pm  0.22$ & C43-4 & $0\farcs70$ & 2018.1.01193.T & K. Maeda\\
58452.1$^{\rm e}$  &    7.1  & 100.0 & $7.42  \pm  0.38$ & C43-4 & $0\farcs94$ & 2018.1.01193.T & K. Maeda\\
58462.1$^{\rm e}$  &   17.1  & 100.0 & $9.05  \pm  0.46$ & C43-4 & $1\farcs17$ & 2018.1.01193.T & K. Maeda\\
58466.0 & 21.0 & 93.5 & $7.41 \pm 0.99$ & C43-4 & 1\farcs14 & 2018.1.01506.S & S. Viti\\
58643.6$^{\rm e}$  & 198.6  & 100.0 & $0.336 \pm  0.026$ & C43-9 & $0\farcs06$ & 2018.A.00038.S & K. Maeda\\
58658.5 & 213.5 & 92.1 & $0.322 \pm 0.036$ & C43-9&0\farcs05&   2018.1.01135.S & J. Wang\\
58751.2 & 306.2 & 103.0 & $0.264 \pm 0.044$ & C43-6& 0\farcs33 &   2018.1.01684.S &  T. Tosaki\\
59469.4 & 1024.4 & 92.1 & $1.128 \pm 0.069$ & C43-9 & 0\farcs05 &   2019.1.00026.S & M. Imanishi\\
59809.4 & 1364.4 & 100.0 & $0.955 \pm 0.093$ & C-5 & 0\farcs76 &  2021.A.00026.S & K. Maeda\\\hline 
{\bf Band 6} & & & & & & & \\\hline
58449.1$^{\rm e}$  &   4.1  & 250.0 & $4.21  \pm  0.43$ & C43-4 & $0\farcs28$ & 2018.1.01193.T & K. Maeda\\
58451.2$^{\rm e}$  &   6.2  & 250.0 & $4.32  \pm  0.44$ & C43-4 & $0\farcs42$ & 2018.1.01193.T & K. Maeda\\
58462.1$^{\rm e}$  &  17.1  & 250.0 & $2.49  \pm  0.28$ & C43-4 & $0\farcs50$ & 2018.1.01193.T & K. Maeda\\
58643.6$^{\rm e}$  & 198.6  & 250.0 & $0.120 \pm  0.022$ & C43-9 & $0\farcs03$ & 2018.A.00038.S & K. Maeda\\
59809.4 & 1364.4 & 250.0 & $0.286 \pm 0.067$ & C-5 & 0\farcs30 &  2021.A.00026.S & K. Maeda\\\hline 
\enddata
\tablecomments{$^{\rm a}$The phase is measured from the putative explosion date (MJD 58445.0). $^{\rm b}$Central frequency. $^{\rm c}$With $1\sigma$ error. $^{\rm d}$Average of the major and minor axes. $^{\rm e}$From \citet{maeda_2018ivc}.
}
\label{tab:flux}
\end{deluxetable*}

The evolution of massive stars toward core collapse and supernova (SN) explosions has not been fully understood, especially in their post main-sequence (MS) evolution \citep[e.g.,][]{langer2012,maeda2022_rev}. Studying the pre-SN mass-loss history, through observational data after the SN explosion, sheds light on this issue; it is essentially the one and only method to probe massive star evolution in the final millennium to even less than one year \citep[e.g.,][]{smith2017hsn}. The radio synchrotron emission, as created at the shock wave formed by the interaction between the SN ejecta and the circumstellar matter (CSM), is a direct probe to the CSM distribution \citep[e.g.,][]{chevalier1998,chevalier2006,maeda2012,matsuoka2019}. For the typical SN shock velocity of $\sim 10,000$ km s$^{-1}$ and the CSM velocity of $\sim 20$ km s$^{-1}$ \citep[for a case of a supergiant progenitor;][]{smith2017hsn}, the nature of the CSM probed by the synchrotron signal at a few years after the explosion translates to the mass-loss rate $\sim 1,000$ years before the explosion. 

SNe IIb are a particularly interesting type of SNe in this context. They are defined by the initial appearance of hydrogen lines in their optical spectra which eventually disappear \citep{filippenko1997}, indicating that they do not have the H-rich envelope as massive as canonical SNe II (red supergiant) while still keeping $\sim 0.01-1 M_\odot$ of the envelope \citep[e.g.,][]{shigeyama1994,woosley1994,bersten2012,hiramatsu2021}, the features which distinguish SNe IIb from the more stripped envelope analogs, i.e., SNe Ib (He star) or SNe Ic (C+O star). As such, SNe IIb bridges SNe II and SNe Ib/c in terms of the pre-SN mass loss mechanism, especially that responsible for ejecting the H-rich envelope \citep[e.g.,][]{fang2019}. The binary evolution channel is a leading scenario for the envelope stripping from SNe II toward SNe IIb/Ib/Ic \citep[e.g.,][]{ouchi2017,yoon2017}, whose progenitors probably share a similar initial mass range \citep[e.g.,][]{fang2019}. Details are, however, yet to be clarified. 

SN 2018ivc, as discovered by the $D < 40$ Mpc SN survey \citep[DLT40;][]{tartaglia2018} on 24 November 2018 (UT) within a day of the explosion \citep{valenti2018,bostroem2020}, is an outlier of the SN IIL subclass, with its faint and rapidly-evolving optical light curve. \citet{maeda_2018ivc} presented and analyzed the data taken by the Atacama Large Millimeter/submillimeter Array (ALMA), covering $4$ to $198$ days since the (well-determined) explosion date, together with the optical and X-ray light curves. They suggested that SN 2018ivc is indeed an analog of an SN IIb explosion (potentially with a slightly more massive H-rich envelope), with its different observational features from canonical SNe IIb originating in its powering mechanism, the SN-CSM interaction. Such a scenario may also be in line with its optical spectral features showing broad and boxy emission lines \citep{dessart2022}. \citet{maeda_2018ivc} further proposed that SN 2018ivc and a fraction of SNe IIL may be a direct link between SNe IIb and SNe IIP in the binary evolution where the sequence of SNe Ic/Ib-IIb-(some) IIL-IIP is mainly controlled by the initial orbital separation. This is along the same lines as previously suggested by \citet{nomoto1995,nomoto1996}, but with a substantial difference in the mode of binary interaction (see Section 4 for more details). 

\begin{figure*}[t]
\centering
\includegraphics[width=1.3\columnwidth]{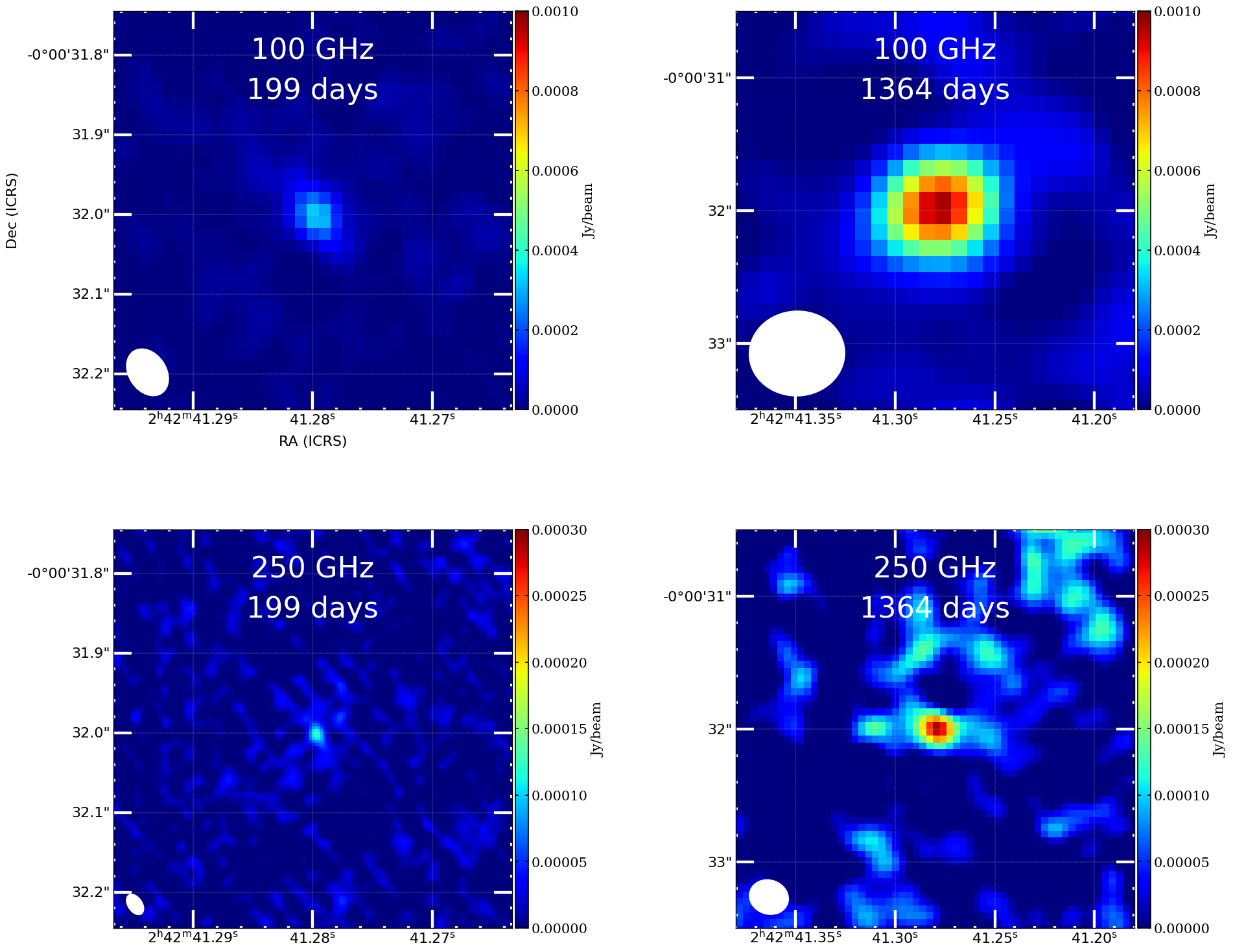}
\caption{The ALMA band 3 (top) and band 6 (bottom) images of SN 2018ivc on day 199 \citep[left;][]{maeda_2018ivc} and day 1,364 (right). The elliptical beam shape is shown on the left-bottom corner in each panel.
}
\label{fig:image}
\end{figure*}

The proximity (8".7 east and 16".1 north) of SN 2018ivc to the core of the well-studied Seyfert galaxy NGC1068 (M77), for which we assume the distance of $10^{+1.8}_{-1.5}$ Mpc \citep{tully2009}, requires a high angular resolution to resolve this SN, but at the same time it offers a unique opportunity for long-term monitoring; the SN location may be covered by archival observations targeting the core of NGC1068. Recently, its detection at $\sim 1,000$ days after the explosion in the {\em cm} emission, at 6.5 GHz by the Karl G. Jansky Very Large Array (VLA) and at 6.3 GHz by the enhanced Multi-Element Radio Linked Interferometer Network (e-MERLIN), was reported \citep{mutie2022}. In this paper, we report the result of our long-term monitoring of SN 2018ivc with ALMA. The paper is structured as follows. In Section 2, we present the observation and date reduction. Results are presented in Section 3, with the analysis of the nature of the CSM at a large scale around SN 2018ivc. Its implications for the progenitor evolution of SN 2018ivc and its relation to other classes of SNe are discussed in Section 4. The paper is summarized in Section 5 with concluding remarks.

\section{Observations and Data Reduction}\label{sec:obs}

In this paper, we report late-time observations of SN 2018ivc ($\gsim 1,000$ days) based on data taken through our ALMA Director's Discretionary Time (DDT) program 2021.A.00026.S (PI: K. Maeda), complemented by the data at earlier epochs published in \citet{maeda_2018ivc} and archival data newly presented in this work. The log of the data used in the present work, together with the measured fluxes, is shown in Table \ref{tab:flux}. The DDT program was conducted on 2022 Aug 18 (1,364 days since the explosion), essentially in the same spectral setup with our early-phase observations up to $\sim 200$ days. On-target exposure time is 5.0 min for band 3 and 9.6 min for band 6. The central frequencies in the continuum bands are 100 and 250\,GHz. The angular resolution is $0\farcs81\times0\farcs71$ in band~3 and $0\farcs33\times0\farcs28$ in band~6 (Figure~1). Since this observation targeted the SN i.e., point-source, the imaging processes are simple. We use the data calibrated through the standard ALMA pipeline with Common Astronomy Software Applications \citep[CASA;][]{casa2022}. The  continuum flux densities associated with the SN on each image are measured based by {\tt imfit} task in {\tt CASA}.

In addition, we use ALMA Band~3 archival data that cover the position of SN 2018ivc. We selected the data in which the SN is detected at the signal-to-noise ratio of $>5$. We downloaded the processed fits continuum images by Japanese Virtual Observatory (JVO). In the cases of 2018.1.01135.S and 2019.1.00026.S, we manually performed the imaging processes, i.e., {\tt clean} task in {\tt casa} with primary beam correction because the archival fits images do not cover the SN position. For 2018.1.01135.S we conducted standard self-calibration to minimize the effects by sidelobes from the bright nuclear emission. The S/N was improved by self-calibration but the flux value associated with the SN is consistent regardless of self-calibration processes. For 2019.1.00026.S, the data were taken during several visits over the course of $\sim$two weeks, and we select those that provide the best S/N ratio at the SN position.

\begin{figure}[t]
\centering
\includegraphics[width=\columnwidth]{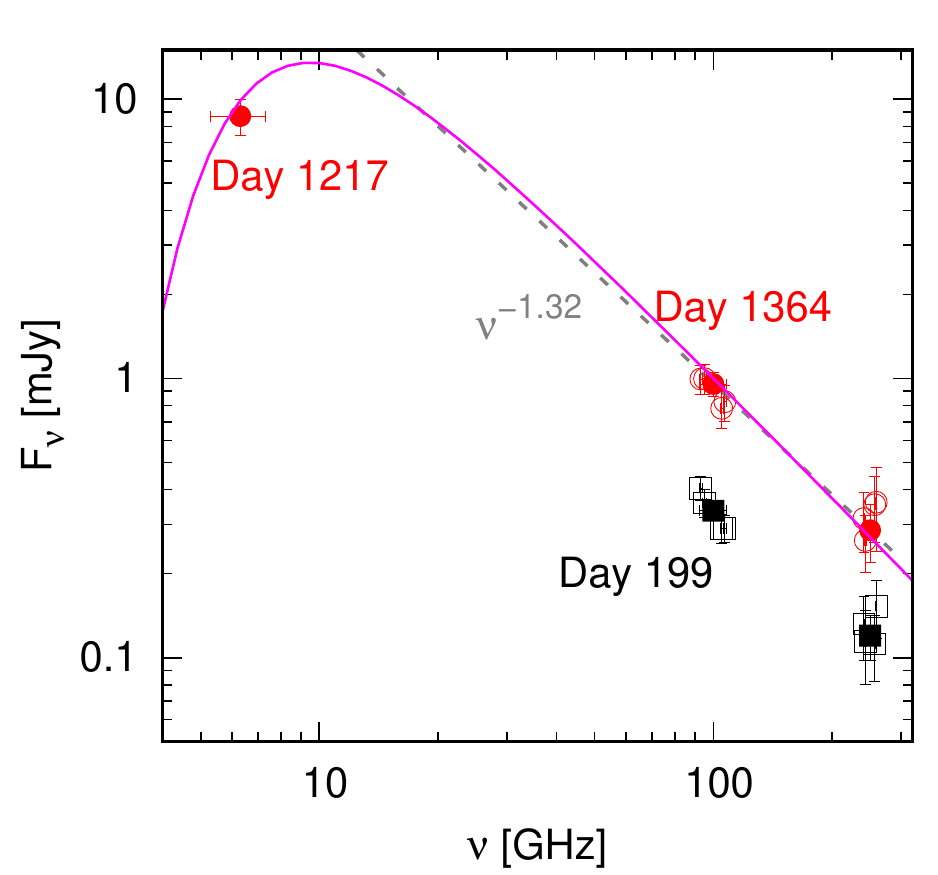}
\caption{The SED of SN 2018ivc on day $\sim 1,300$ (red-circles), as constructed from the data at 6.3 GHz \citep{mutie2022} and at 100/250 GHz (this work). For comparison, the SED on day 199 (black-squares) is also shown \citep{maeda_2018ivc}. The flux densities derived for the individual spectral windows (SPWs) are shown by open symbols, while the flux densities after combining the four continuum spectral windows within each band are shown by filled symbols. The flux densities are shown with $1\sigma$ error; the flux calibration error is included for the SPW-combined data, while it is omitted for the individual SPW data. The power-law fit to the ALMA data on day 1364 is shown by the dashed line. For demonstration purposes, the SED computed for the model with $s=1$ is shown (magenta; see also Fig. \ref{fig:lc}). 
}
\label{fig:sed}
\end{figure}

\begin{figure*}[t]
\centering
\includegraphics[width=\columnwidth]{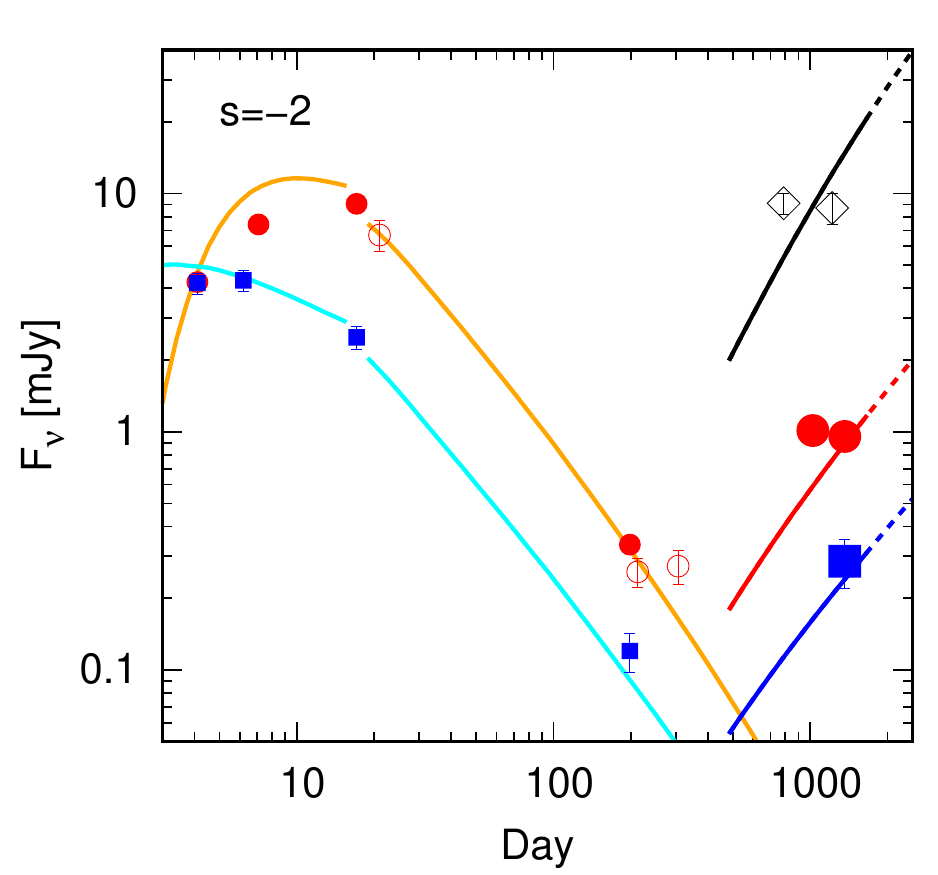}
\includegraphics[width=\columnwidth]{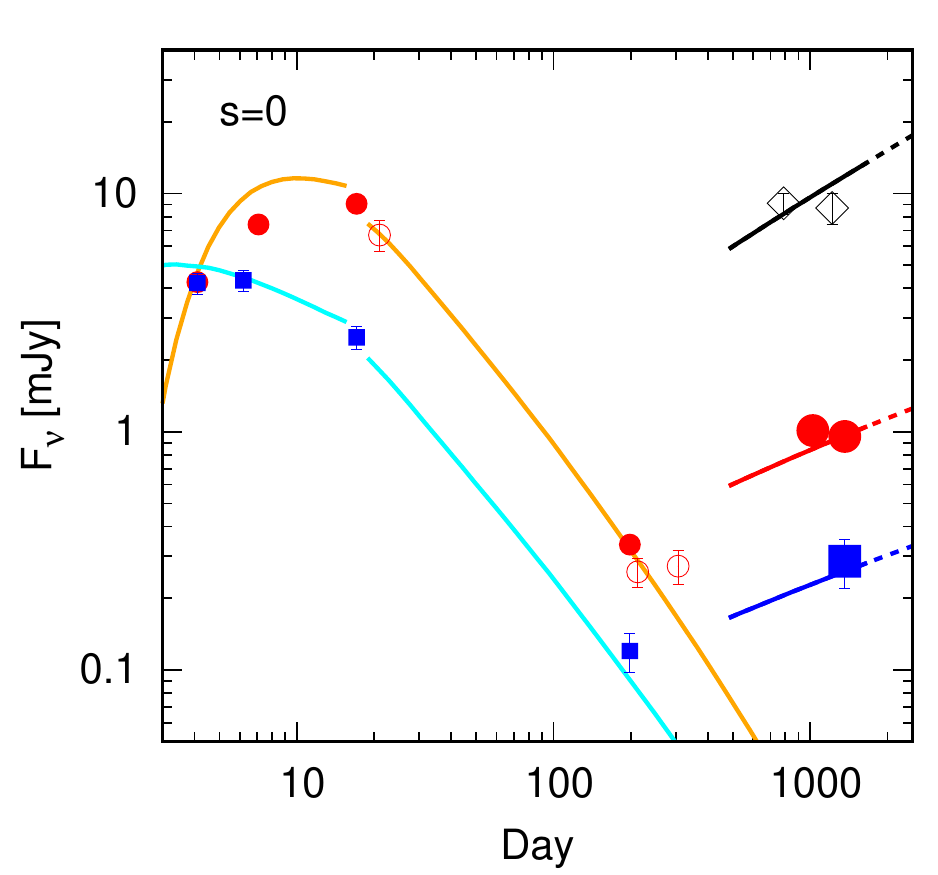}
\includegraphics[width=\columnwidth]{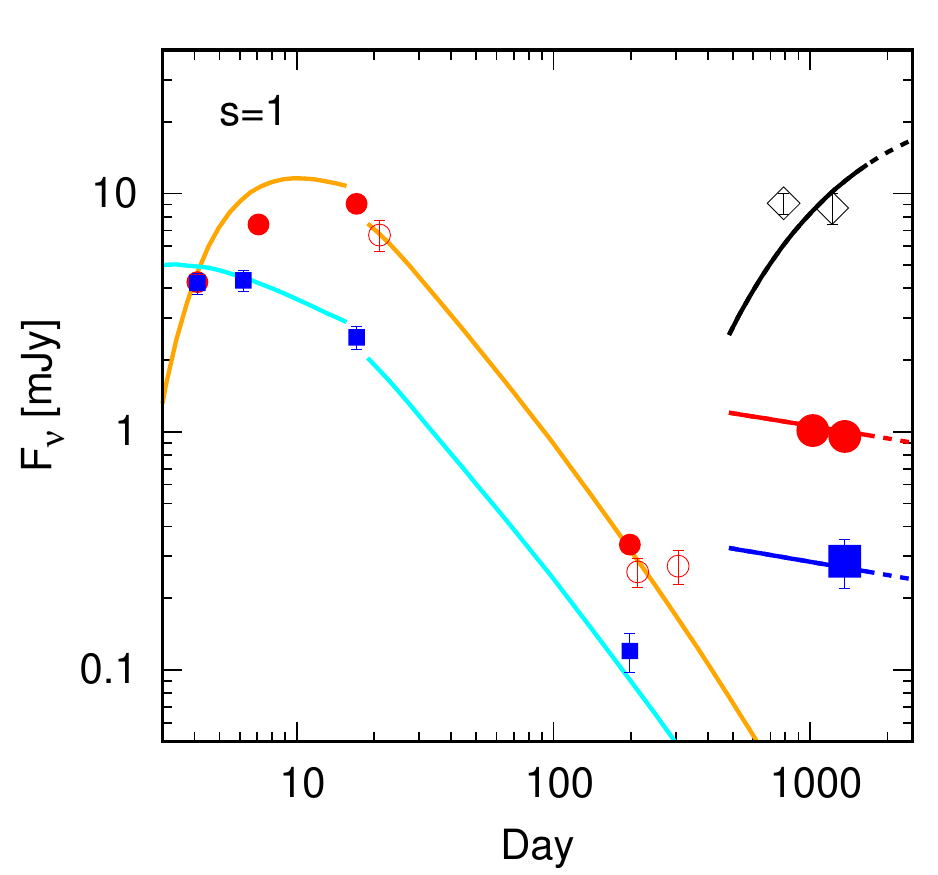}
\includegraphics[width=\columnwidth]{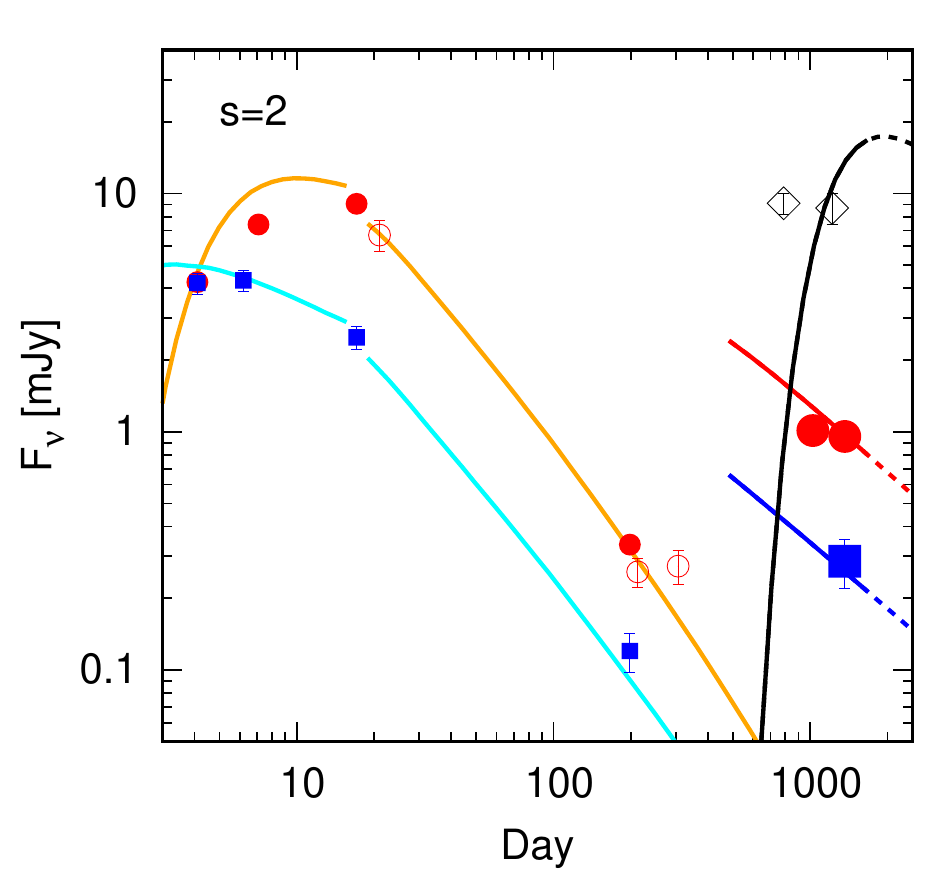}
\caption{The radio light curves of SN 2018ivc. The high-frequency data are taken by the ALMA, at 100 GHz (red circles) and 250 GHz (blue squares), including the data from \citet{maeda_2018ivc} (small filled symbols), the new late-time data at $> 1,000$ days (large filled), and the other archival data (open symbols). The late-time {\em cm} flux densities at 6.3/6.5 GHz (black open diamonds) are taken from \citet{mutie2022}. For band 3 data, the flux densities are corrected to those at 100 GHz, assuming the SED slope seen in a neighboring epoch when the SN was observed both in bands 3 and 6 (see Fig. \ref{fig:sed}). The model curves at 100 (orange) and 250GHz (cyan) in the earlier phase ($\lsim 200$ days) are from \citet{maeda_2018ivc}. Four panels here are shown with different models for the late-phase light curves, at 6.5 GHz (black), 100 GHz (red), and 250 GHz (blue) with $s=-2, 0, 1, 2$ for the CSM slope. 
}
\label{fig:lc}
\end{figure*}

\section{Results}\label{sec:results}

Fig. \ref{fig:image} shows the reconstructed images of SN 2018ivc on day 199 \citep{maeda_2018ivc} and day 1364 (this work). After the flux decay in the earlier phase up to $\sim 200$ days, the SN has begun brightening again. There is a hint of a slow decay between days 1,024 and 1,364, but they are also consistent with the same flux level within $1\sigma$ (after correcting for the central frequency difference). 

Fig. \ref{fig:sed} compares the Spectral Energy Distribution (SED) on days 199 and 1,364 , with the addition of the {\em cm} flux on day 1,217 taken from \citet{mutie2022}. It is clear that SN 2018ivc remains optically thin in the high frequency emission in the ALMA bands. Assuming $f_{\nu} \propto \nu^{\alpha}$, $\alpha = -1.32 \pm 0.15  (1\sigma)$ is measured on day 1,364. Given the SED slope of $\alpha = -1.12 \pm  0.22$ on day 199 \citep{maeda_2018ivc}, there is a hint of the spectral steepening but it is also consistent with no change within $1\sigma$ (i.e., $\alpha = -1.22 \pm 0.27$ for the combined analysis of the data on days 199 and 1,364). The SED as found in our ALMA observations also indicates that the {\em cm} emission must be in the optically-thick regime. This highlights the power of the higher-frequency observation despite generally the lower flux level; catching the optically-thin portion of the synchrotron emission is essential to constrain the CSM properties \citep{maeda2021,maeda_2018ivc}. 

Fig. \ref{fig:lc} shows the light curves. Rebrightening at 100 and 250 GHz is clearly seen, which probably started at $\sim 300$ days. Since it has been optically thin in the ALMA bands both before and after the rebrightening (Fig. \ref{fig:sed}), the flux increase means that the CSM is not smoothly distributed, and that the CSM density in the outer region must be higher than the extrapolation from the inner region. 

We have computed a series of synchrotron emission models \citep[see][]{maeda2021,maeda_2018ivc}, by considering a CSM extended outward from $\sim 4 \times 10^{16}$ cm (i.e., with the main interaction taking place at $\sim 500$ days for the SN ejecta velocity of $\sim 10,000$ km s$^{-1}$). In practice, we simply assume the CSM is distributed with the density structure of $\rho_{\rm CSM} \propto D r^{-s}$ without a hole in the computation (where $D = \dot M / 4 \pi v_{\rm w}$, $\dot M$ is the mass-loss rate, and $v_{\rm w}$ is the mass-loss wind velocity); once the interaction has fully developed at $\sim 500$ days (at $\sim 4 \times 10^{16}$ cm), the mass of the swept up CSM is expected to be soon dominated by the outer component, and therefore the evolution of the SN-CSM interaction will quickly lose the memory of the earlier interaction and follow the asymptotic behavior determined by the outer CSM component. We thus apply this model to the epoch after $\sim 500$ days. 

The ejecta structure and the microphysics parameters are the same with those adopted by \citet{maeda_2018ivc} for the early-phase modeling; $M_{\rm ej} = 3 M_\odot$, $E_{\rm K} = 1.2 \times 10^{51}$ erg, and the outer ejeca density slope of $n=-10$. The {\em cm} emission is in the optically-thick regime, and we find that the synchrotron self-absorption alone would not explain the cut off at the low frequency. We thus include free-free absorption which is however poorly constrained by theory. We simply vary the optical depth by changing the constant electron temperature in the unshocked CSM. It is just for demonstration, and in most of the subsequent analyses we do not use constraints from the {\em cm} observations due to the large uncertainty in the model framework. 

\begin{figure*}[t]
\centering
\includegraphics[width=\columnwidth]{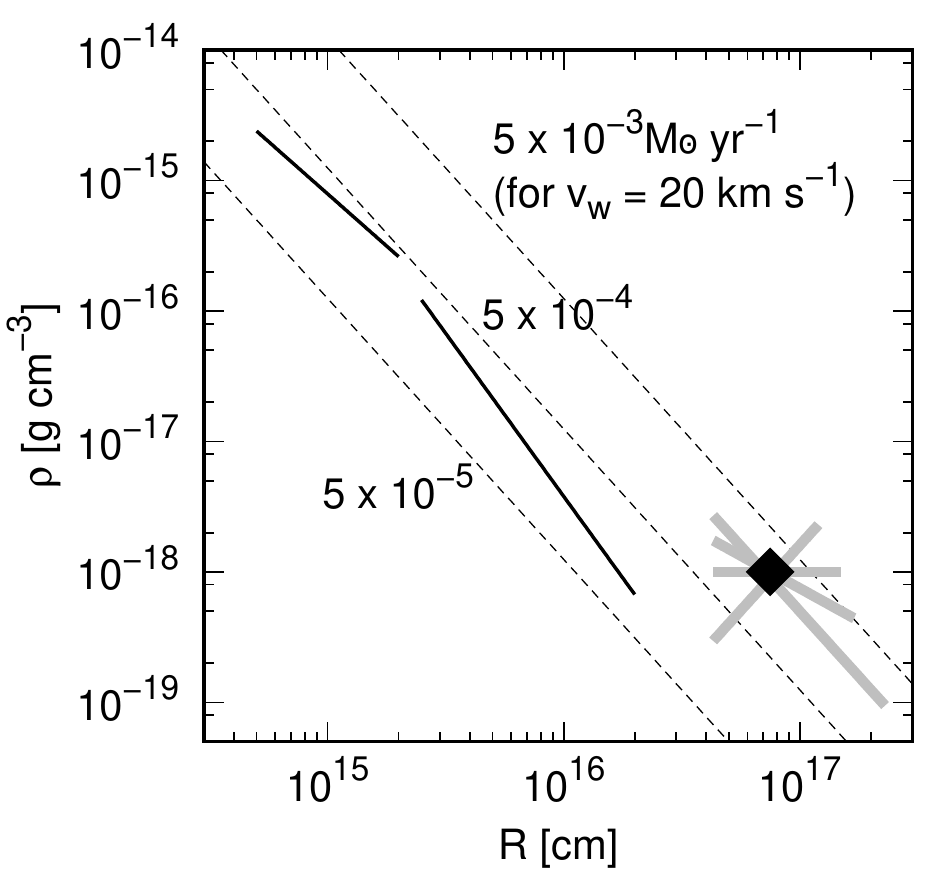}
\includegraphics[width=\columnwidth]{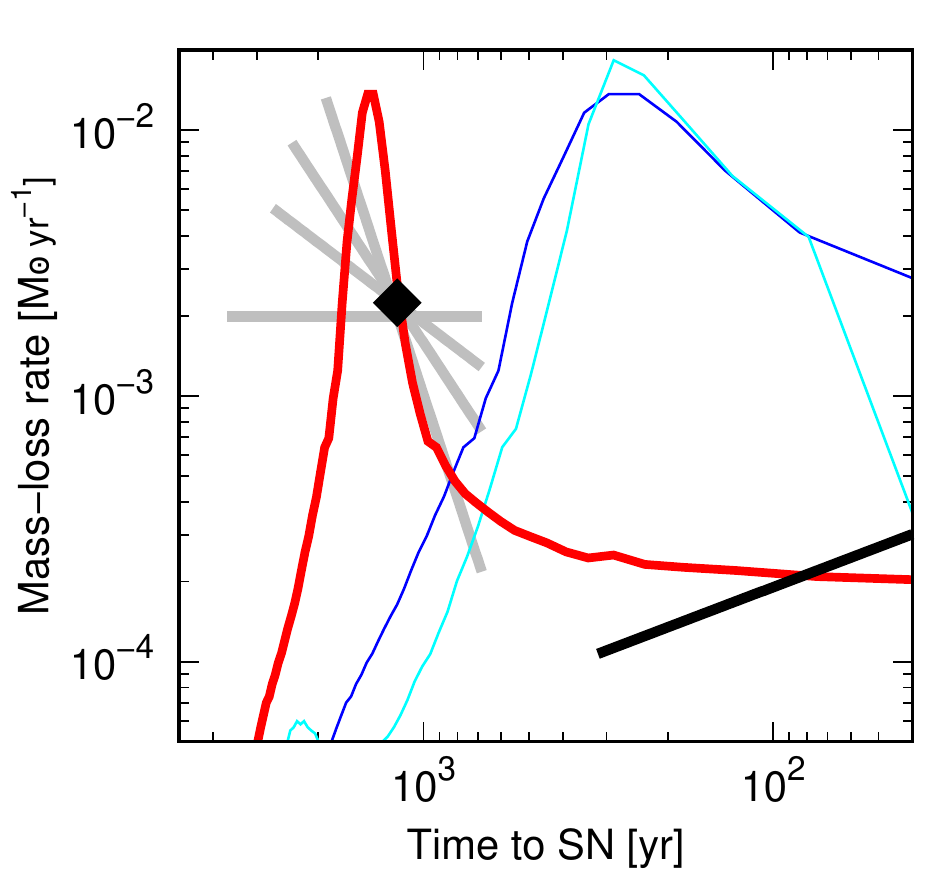}
\caption{\textit{Left:} The CSM distribution derived for SN 2018ivc. The inner structure at $\lsim 2 \times 10^{16}$ cm (black lines) is from \citet{maeda_2018ivc}. The CSM distributions adopted in the light curve models ($s=-2, 0, 1, 2$) for the late epochs (Fig. \ref{fig:lc}) are shown by the gray lines, where the overlapping region (black diamond) is strongly constrained irrespective of the model details. \textit{Right:} The mass-loss history toward the SN (black line and point for SN 2018ivc). The mass-loss histories based on Models 25 of \citet{ouchi2017} (peak at 5,800 yr before the SN) and 28 (peak at $\sim 300$ yr) are shown for comparison; Model 28 (cyan), Model 25 shifted toward the SN by 5,450 yr (blue) and by 4,300 yr (red). 
}
\label{fig:csm}
\end{figure*}

Fig. \ref{fig:lc} shows models with four different CSM distributions; $s=-2$ (i.e., increasing density), $0$ (constant), $1$, and $2$ (steady-state mass loss). These CSM distributions are plotted in Fig. \ref{fig:csm}. By changing the density scale ($D$), all the models can reproduce the flux levels at the latest epochs (1,024 and 1,364 days). The flat ($s=0$) and increasing ($s=-2$) CSM distributions are not favored, given the nearly flat or slowly decaying evolution observed during this time window. The model with $s=1$ provides the best representation of the evolution, while our focus in the present work is the CSM density scale without aiming at constraining the value of $s$. It is possible to explain the flux levels at 6.5/6.3 GHz in the optically-thick emission; within a simple treatment of the free-free absorption we are however not able to explain the flat or slowly-decaying evolution also seen in the {\em cm} emission. We emphasize that the {\em cm} emission, including both the model and the calibration of the data, is not a topic of this work. 

While there is a degeneracy in the CSM slope given the limited temporal coverage, the CSM density at $\sim 10^{17}$ cm is strongly constrained to be $\sim 10^{-18}$ g cm$^{-3}$; this is where the data points exist. It is substantially above the extrapolation from the inner structure, and it is about an order of magnitude larger in terms of the mass-loss rate (assuming that $v_{\rm w}$ did not change). 

\section{Discussion}\label{sec:discussion}
In the SN `IIb' scenario with an extended progenitor \citep{maeda_2018ivc}, the mass-loss velocity would be $\sim 20$ km s$^{-1}$ \citep[e.g.,][]{groh2014,smith2017hsn}. Then, the location of the `outer' dense CSM ($\sim 10^{17}$ cm) corresponds to the mass-loss history $\sim 1,500$ yr before the explosion. Based on the binary evolution model toward SNe IIb \citep{ouchi2017},  \citet{maeda_2018ivc} suggested that the progenitor of SN 2018ivc might have experienced an extreme case of the Case C mass transfer, where the binary mass transfer is initiated in the advanced evolution stage after the core He ignition \citep[e.g.,][]{langer2012}, to explain the high-mass loss rate just before the explosion as inferred from the early-phase observation \citep[see also][]{maeda2015}. As compared to other SNe IIb, it is suggested that the strong binary interaction has taken place later in its evolution, closer to the time of the core-collapse explosion, which is attributed to larger initial orbital separation. 

\citet{ouchi2017} indeed predicted that the extreme case C mass transfer may bridge the evolution toward SNe IIb and SNe IIP, where the SN as surrounded by a dense CSM will be identified either as SNe IIL or IIn. In Fig. \ref{fig:csm}, we show two such models showing late-time binary interaction \citep{ouchi2017}, for the initial masses of the primary and the secondary of $16 M_\odot$ and $9.6 M_\odot$, with the only difference in its initial orbital period (1,800 day for their Model 25, and 1,950 day for Model 28). In both models, the binary first experiences rapid mass transfer until the mass ratio is inverted, during which most of the H-rich envelope ($\gsim 5 M_\odot$) is ejected. This is then followed by a relatively high mass-transfer rate of $\gsim 10^{-5} M_\odot$ yr$^{-1}$ as the primary keeps filling the Roche Lobe (RL). The final left-over envelope mass is $\sim 2 M_\odot$ in these models, which is somewhat large as an `SN IIb' but qualitatively consistent with the extended `SN IIb' progenitor scenario with slightly more massive envelope than canonical SNe IIb; the final envelope mass should be dependent on the binary parameters, especially on the mass ratio that roughly controls the mass of the left-over envelope after the rapid transfer phase (see above). The initial rapid transfer takes place $\sim 5,800$ yr and $\sim 300$ yr before the explosion, for Models 25 and 28, respectively. 

While the timing is different, the mass-loss properties during the initial rapid-transfer phase are similar (Fig. \ref{fig:csm}). It is thus in principle possible to have the strong interaction phase at $\sim 1,500$ yr where most of the envelope, $\gsim 5 M_\odot$, is ejected, which form the dense CSM at $\sim 10^{17}$ cm, by a different combination of the component masses and the initial separation. This is demonstrated in Fig. \ref{fig:csm} by comparing the mass-loss history derived for SN 2018ivc and those predicted by Models 25 and 28 of \citet{ouchi2017}. Indeed, if we shift the mass-loss history of Model 25 toward the explosion date, assuming slightly different binary parameters (e.g., the initial separation and/or the initial mass ratio), it can explain that derived for SN 2018ivc reasonably well. 

In optical wavelengths, SNe IIb from `extended' progenitors tend to show signs of the SN-CSM interaction in the late phase ($\gsim 1$ yr), as the emergence of strong H$_\alpha$ emission and light curve flattening \citep[][]{patat1995,matheson2000,maeda2015,fremling2019}. However, it is more likely that the late-time emergence of the SN-CSM interaction in these cases is simply due to the decreasing importance of other powering mechanisms (e.g., $^{56}$Ni/Co decay) and it would not require an increase of the SN-CSM interaction power by a distinct high-density CSM component \citep{maeda2015}. It is naturally expected if the timing of the strong binary interaction occurred earlier for these systems than for SN 2018ivc; in this case, the CSM created by the first, rapid mass-transfer phase must be located far outside. 

To our knowledge, this is the second `SN IIb' for which a clear radio rebrightening has been detected; the first example is the somewhat peculiar SN IIb 2003bg, suggested to be a broad-lined SN IIb \citep{hamuy2009} that showed a jump in the radio flux by a factor of two or three at $\sim 100$ days \citep{soderberg2006}. We also note that SN IIb 2001ig showed a modulation superimposed on a long-term decline in its radio light curves \citep{ryder2004}. These SNe IIb may share some common evolutionary path to SN 2018ivc. Possible modulations were also seen in SN IIb 2004C \citep{demarchi2022}, though the sparse sampling at such late times makes it hard to ascertain their magnitude and timescale. In any event SN 2004C did not show a sustained transition from decay to rebrightening. SN 2004C is nevertheless one of the only two SNe IIb included in the sample of late-time radio monitoring by \citet{stroh2021}, with the other example lacking information on its temporal evolution. 

Radio rebrightening is rare for SNe, and only a handful of events are known so far \citep[e.g.,][]{bauer2008}. Of particular interests are a few SNe Ib/c showing transition to SNe IIn and the late-time radio rebrightening with the time scale covering $\sim$ a year to decades \citep{anderson2017,chandra2020,thomas2022}; \citet{margutti2017} showed that about 10\% of SNe Ib/c exhibit radio ({\em cm}) rebrightening at $\sim 1$ yr after the explosion with typically an order of magnitude enhancement in the flux. It should be interesting to compare the characteristics of SN 2018ivc with those events, especially SN 2014C as the most well-studied example. \citet{margutti2017} argued that the inner region is more like a cavity, surround by an extremely dense CSM; extremely high-mass loss rate ($\sim 1 M_\odot$ yr$^{-1}$) at $\sim 10-1,000$ years before the explosion was followed by the mass loss consistent with a wind from a Wolf-Rayet star \citep[$\lsim 10^{-5} M_\odot$ yr$^{-1}$][]{smith2017hsn}. \citet{margutti2017} suggested as one possibility the scenario where the common envelope is responsible for the creation of the detached dense CSM, leaving a bare He star. The case for SN 2018ivc has two remarkable differences: (1) the initial, main mass loss ($\sim 1,500$ yr before the explosion) is less dramatic, and then (2) it keeps a relatively high mass-loss rate toward the explosion. Therefore, it fits better to the scenario proposed here \citep[see also][]{ouchi2017,maeda_2018ivc}; the initial mass transfer was rapid but did not enter into the common envelope, and the rapid phase was over when the mass ratio was inverted. At that moment $\sim 1 M_\odot$ of the H-rich envelope was left, which keeps filling the RL. An interesting, alternative scenario may be a merger of the cores following the common envelope, in which the excited merger product may keep a high mass loss rate up until the explosion \citep{nomoto1995,nomoto1996}. 

The {\em mm} observations of SNe are extremely rare, therefore virtually no sample is available to make any data-based estimate on the frequency of SNe showing rebrightening at {\em mm} wavelengths. Indeed, a well-defined control sample is still missing even at {\em cm} wavelengths \citep{bietenholz2021}, which makes even a qualitative estimate of the frequency of rebrightening events somewhat fraught, with the possible exception of SNe Ib/c \citep[see above:][]{margutti2017}. Alternatively, we could estimate an expected rate of such events based on the scenario proposed here (i.e., the boundary between the single and binary evolution channels). \citet{maeda_2018ivc} gave such an estimate for the `SN 2018ivc-like' events based on the model sequence of \citet{ouchi2017}, as being $\sim 10$\%  of SNe IIb and $\sim 3-6$\% of SESNe. This estimate indeed indicates that a non-negligible fraction of SNe Ib/c showing rebrightening ($\sim 10\%$ of SNe Ib/c; see above) may be linked to this binary evolution channel. On the other hand, the same fraction was estimated to be $\sim 10-20$\% of SNe IIL, indicating that the evolutionary channel considered here is not a major channel toward SNe IIL, which might reflect potential `mixed populations' within SNe IIL (\citealt{arcavi2012}; but see also \citealt{anderson2014}). As compared to SNe IIP, SN 2018ivc-like events are very rare, with the expected rate of only $\sim 3$\% of SNe IIP (with large uncertainties) where we assume 
that the relative proportion of SNe IIP to SNe IIL is $\sim$ 7:1 \citep{li2011}. These crude estimates suggest that the SN 2018ivc-like events, or the events linking the single and binary evolution channels, are intrinsically rare in the population of H-rich SNe. SNe IIL would make especially interesting targets, as the above estimate suggests that the bulk of SNe IIL are unlikely to have originated in the specific binary evolutionary channel proposed here. A future sample of SNe IIL followed up in the radio, particularly at {\em mm} wavelengths, may reveal whether such rebrightening is exclusively associated with the SN2018ivc-like objects. 

Another possibility for the detached CSM is an external effect. For the detached CSM shell associated with SN 2014C, \citet{milisavljevic2015} considered photoionization confinement of the CSM by an external radiation field imposed by neighboring stars in a stellar cluster, based on the scenario proposed by \citet{mackey2014}. This scenario, however, is unlikely to work for SN 2018ivc; \citet{bostroem2020} placed an upper limit of the MS mass of a `single star' being 
$\lsim 11 M_\odot$ that existed at the SN location in pre-SN Hubble Space Telescope images. 

\section{Concluding Remarks}\label{sec:concluding}

We have presented long-term monitoring of a peculiar SN IIL 2018ivc ($\sim 4 - 1,364$ days) with ALMA. SN 2018ivc started showing rebrightening in its synchrotron flux at $\sim 300 - 500$ days after the explosion. The radial CSM distribution as reconstructed from the long-term {\em mm} light curves shows a very high mass-loss rate exceeding $10^{-3} M_\odot$ yr$^{-1}$ at $\sim 1,500$ years before the explosion, followed by a moderately high mass-loss rate of $\gsim 10^{-4} M_\odot$ yr$^{-1}$ in the remaining evolution toward the SN explosion. This behavior is in line with the expectation from the scenario in which SN 2018ivc is indeed more like an `extended SN IIb' in its ejecta properties (i.e., He star surrounded by $\sim 1 M_\odot$ of the H-rich envelope) produced through the binary with a relatively large initial separation at a boundary between SNe IIb and SNe IIP \citep{ouchi2017,maeda_2018ivc}.

To our knowledge, it is the first example where rebrightening is detected in the {\em mm} synchrotron emission from SNe. Further, in the above-mentioned scenario, it is the second example where the radio-synchrotron rebrightening has been observed for an `SN IIb'. Indeed, SNe showing radio rebrightening are very rare. The comparison to the properties of well-observed SN 2014C, showing a transition from SN Ib to SN IIn, shows that they likely share the similar (binary) evolution channel albeit with some different details. It is likely that the strong binary interaction started at similar epochs ($\sim 100 - 1,000$ yrs before the explosion), while the outcome is different; SN 2018ivc is more likely explained by the rapid RL over-flow (non-conservative) mass transfer leaving a RL-filling primary star, while SN 2014C entered into the common envelope and stripped away essentially all the H-rich envelope. Identifying the cause of the difference is beyond the scope of the present work, but it may be for example due to the masses of the binary components (e.g., the mass ratio). 

The models presented here suggest that $\sim 1 M_\odot$ of the CSM has been swept up by day $\sim 1,500$. As this is comparable to the expected envelope mass, the shock will start experiencing rapid deceleration afterward. Therefore, we predict that the synchrotron flux will shortly turn into the decay phase again. Further, if the mass budget for the CSM is $\sim 5-10 M_\odot$ as expected in the binary evolution scenario for SNe IIb (and SESNe), we may assume that the characteristic shock velocity would be $\sim 5,000$ km s$^{-1}$ after a substantial amount of this CSM is swept up. Given that there is additional CSM for another $(0.5-1) \times 10^{17}$ cm, the shock will eventually break out of this CSM in $3-6$ years after which the interaction will be essentially over. As such, we plan to continuously monitor SN 2018ivc, which could be done in part of the monitoring observations of NGC1068. In addition, it can be an interesting target for a continuous monitoring with Very Long Baseline Interferometry; we expect to start spatially resolving the shock expansion at $\gsim 5$ yrs for the angular resolution of $\sim 0\farcs001$. It can be an interesting target not only for existing facilities, but for future facilities such as the next-generation Very Large Array \citep[ngVLA:][]{murphy2018} . 

\acknowledgments

The authors thank the referee for constructive comments. This paper makes use of the following ALMA data: ADS/JAO.ALMA \#2018.1.01135.S, \#2018.1.01193.T, \#2018.1.01506.S,  \#2018.1.01684.S, \#2018.A.00038.S., \#2019.1.00026.S, and \#2021.A.00026.S. Some of the ALMA data were retrieved from the JVO portal (\url{http://jvo.nao.ac.jp/portal/}) operated by ADC/NAOJ.
ALMA is a partnership of ESO (representing its member states), NSF (USA) and NINS (Japan), together with NRC (Canada), MOST and ASIAA (Taiwan), and KASI (Republic of Korea), in cooperation with the Republic of Chile. The Joint ALMA Observatory is operated by ESO, AUI/NRAO and NAOJ. K.M. acknowledges support from the Japan Society for the Promotion of Science (JSPS) KAKENHI grant JP18H05223,  JP20H00174, and JP20H04737. K.M. was supported by the ALMA Japan Research Grant of NAOJ ALMA Project, NAOJ-ALMA-269. T.M. appreciates the support from NAOJ ALMA Scientific Research grant No. 2021-17A. T.M. is supported by JSPS KAKENHI grant No. JP22K14073. H.K. was funded by the Academy of Finland projects 324504 and 328898. M.I. acknowledges support from JSPS KAKENHI grant JP21K03632. The work is partly supported by the JSPS Open Partnership Bilateral Joint Research Projects between Japan and Finland (K.M and H.K; JPJSBP120229923). 

%








\bibliography{sn2018ivc_late}{}
\bibliographystyle{aasjournal}



\end{document}